%Paper: hep-th/9306164
%From: jmf@avzw02.physik.uni-bonn.de (Jose Miguel Figueroa-O'Farrill)
%Date: Wed, 30 Jun 93 20:17:20 +0200
%Date (revised): Thu, 1 Jul 93 14:34:12 +0200
%Date (revised): Thu, 1 Jul 93 21:58:07 +0200
%Date (revised): Sun, 4 Jul 93 22:33:45 +0200

%%%%%%%%%%%%%%%%%%%%%%%%%%%%%%%%%%%%%%%%%%%%%%%%%%%%%%%%%%%%%%%%
%
%   This is the Plain TeX file for
%
%
%         Affine Algebras, $N{=}2$ Superconformal Algebras,
%
%                     and Gauged WZNW Models
%
%       by
%
%                   J.M. Figueroa-O'Farrill
%
% ( If you have problems compiling the TeX file please contact me at
%
%               jmf@avzw01.physik.uni-bonn.de  )
%
%%%%%%%%%%%%%%%%%%%%%%%%%%%%%%%%%%%%%%%%%%%%%%%%%%%%%%%%%%%%%
%%%These are the macros for submission of papers to hep-th%%%
%%%The default setting is 12pt and 1 page/side but in the%%%%
%%%future it may allow people to choose also 10 pt and%%%%%%%
%%%2 pages/side.%%%%%%%%%%%%%%%%%%%%%%%%%%%%%%%%%%%%%%%%%%%%%
%%%%%%%%%%%%%%%%%%%%%%%%%%%%%%%%%%%%%%%%%%%%%%%%%%%%%%%%%%%%%
%
\def\unlockat{\catcode`\@=11}
\def\lockat{\catcode`\@=12}
\unlockat
\def\d@f@ult{} \newif\ifamsfonts \newif\ifafour
%
% \def\m@ssage{\immediate\write16}  \m@ssage{}
% \m@ssage{hep-th preprint macros.  Last modified 16/10/92 (jmf).}
% \message{These macros work with AMS Fonts 2.1 (available via ftp from}
% \message{e-math.ams.com).  If you have them simply hit "return"; if}
% \message{you don't, type "n" now: }
% \endlinechar=-1  %don't add spaces at end of line
% \read-1 to\@nswer
% \endlinechar=13
% \ifx\@nswer\d@f@ult\amsfontstrue
%     \m@ssage{(Will load AMS fonts.)}
% \else\amsfontsfalse\m@ssage{(Won't load AMS fonts.)}\fi
% %
% \message{The default papersize is A4.  If you use US 8.5" x 11"}
% \message{type an "a" now, else just hit "return": }
% \endlinechar=-1  %don't add spaces at end of line
% \read-1 to\@nswer
% \endlinechar=13
% \ifx\@nswer\d@f@ult\afourtrue
%     \m@ssage{(Using A4 paper.)}
% \else\afourfalse\m@ssage{(Using US 8.5" x 11".)}\fi
% %
% \nonstopmode
%
%%%%%%%%%%%%%%%%%%%%%%
%%%Font definitions%%%
%%%%%%%%%%%%%%%%%%%%%%
%

\font\twelverm=cmr12
\font\ninerm=cmr9
\font\sixrm=cmr6
\font\fourteenbf=cmbx12 scaled\magstep1
\font\twelvebf=cmbx12
\font\ninebf=cmbx9
\font\sixbf=cmbx6
\font\fourteeni=cmmi12 scaled\magstep1      \skewchar\fourteeni='177
\font\twelvei=cmmi12                        \skewchar\twelvei='177
\font\ninei=cmmi9                           \skewchar\ninei='177
\font\sixi=cmmi6                            \skewchar\sixi='177
\font\fourteensy=cmsy10 scaled\magstep2     \skewchar\fourteensy='60
\font\twelvesy=cmsy10 scaled\magstep1       \skewchar\twelvesy='60
\font\ninesy=cmsy9                          \skewchar\ninesy='60
\font\sixsy=cmsy6                           \skewchar\sixsy='60
\font\fourteenex=cmex10 scaled\magstep2
\font\twelveex=cmex10 scaled\magstep1

\ifamsfonts
   \font\ninex=cmex9
   
   \font\sixex=cmex7 at 6pt
   
\else
   \font\ninex=cmex10 at 9pt
   
   \font\sixex=cmex10 at 6pt
   
\fi
\font\fourteensl=cmsl10 scaled\magstep2
\font\twelvesl=cmsl10 scaled\magstep1

\font\sevensl=cmsl10 at 7pt
\font\sixsl=cmsl10 at 6pt

\font\fourteenit=cmti12 scaled\magstep1
\font\twelveit=cmti12

\font\fourteentt=cmtt12 scaled\magstep1
\font\twelvett=cmtt12
\font\fourteencp=cmcsc10 scaled\magstep2
\font\twelvecp=cmcsc10 scaled\magstep1

\ifamsfonts
   
\else
   
\fi
\newfam\cpfam
\font\fourteenss=cmss12 scaled\magstep1
\font\twelvess=cmss12
\font\tenss=cmss10
\font\niness=cmss9

\font\sevenss=cmss8 at 7pt
\font\sixss=cmss8 at 6pt
\newfam\ssfam
\newfam\msafam \newfam\msbfam \newfam\eufam
\ifamsfonts
 \font\fourteenmsa=msam10 scaled\magstep2
 \font\twelvemsa=msam10 scaled\magstep1
 \font\tenmsa=msam10
 \font\ninemsa=msam9
 \font\sevenmsa=msam7
 \font\sixmsa=msam6
 \font\fourteenmsb=msbm10 scaled\magstep2
 \font\twelvemsb=msbm10 scaled\magstep1
 \font\tenmsb=msbm10
 \font\ninemsb=msbm9
 \font\sevenmsb=msbm7
 \font\sixmsb=msbm6
 \font\fourteeneu=eufm10 scaled\magstep2
 \font\twelveeu=eufm10 scaled\magstep1
 \font\teneu=eufm10
 \font\nineeu=eufm9
 
 \font\seveneu=eufm7
 \font\sixeu=eufm6
 \def\hexnumber@#1{\ifnum#1<10 \number#1\else
  \ifnum#1=10 A\else\ifnum#1=11 B\else\ifnum#1=12 C\else
  \ifnum#1=13 D\else\ifnum#1=14 E\else\ifnum#1=15 F\fi\fi\fi\fi\fi\fi\fi}
 \def\hexmsa{\hexnumber@\msafam}
 \def\hexmsb{\hexnumber@\msbfam} 
\fi
\newdimen\b@gheight             \b@gheight=12pt
\newcount\f@ntkey               \f@ntkey=0
\def\f@m{\afterassignment\samef@nt\f@ntkey=}
\def\samef@nt{\fam=\f@ntkey \the\textfont\f@ntkey\relax}
\def\rm{\f@m0 }
\def\mit{\f@m1 }
\def\cal{\f@m2 }
\def\it{\f@m\itfam}
\def\sl{\f@m\slfam}
\def\bf{\f@m\bffam}
\def\tt{\f@m\ttfam}
\def\caps{\f@m\cpfam}
\def\ssf{\f@m\ssfam}
\ifamsfonts
 \def\msa{\f@m\msafam}
 \def\msb{\f@m\msbfam} 
 \def\eu{\f@m\eufam}
\else
  \let\eu=\bf
\fi
\def\fourteenpoint{\relax
    \textfont0=\fourteencp          \scriptfont0=\tenrm
      \scriptscriptfont0=\sevenrm
    \textfont1=\fourteeni           \scriptfont1=\teni
      \scriptscriptfont1=\seveni
    \textfont2=\fourteensy          \scriptfont2=\tensy
      \scriptscriptfont2=\sevensy
    \textfont3=\fourteenex          \scriptfont3=\twelveex
      \scriptscriptfont3=\tenex
    \textfont\itfam=\fourteenit     \scriptfont\itfam=\tenit
    \textfont\slfam=\fourteensl     \scriptfont\slfam=\tensl
      \scriptscriptfont\slfam=\sevensl
    \textfont\bffam=\fourteenbf     \scriptfont\bffam=\tenbf
      \scriptscriptfont\bffam=\sevenbf
    \textfont\ttfam=\fourteentt
    \textfont\cpfam=\fourteencp
    \textfont\ssfam=\fourteenss     \scriptfont\ssfam=\tenss
      \scriptscriptfont\ssfam=\sevenss
    \ifamsfonts
       \textfont\msafam=\fourteenmsa     \scriptfont\msafam=\tenmsa
         \scriptscriptfont\msafam=\sevenmsa
       \textfont\msbfam=\fourteenmsb     \scriptfont\msbfam=\tenmsb
         \scriptscriptfont\msbfam=\sevenmsb
       \textfont\eufam=\fourteeneu     \scriptfont\eufam=\teneu
         \scriptscriptfont\eufam=\seveneu \fi
    \samef@nt
    \b@gheight=14pt
    \setbox\strutbox=\hbox{\vrule height 0.85\b@gheight
                                depth 0.35\b@gheight width\z@ }}
\def\twelvepoint{\relax
    \textfont0=\twelverm          \scriptfont0=\ninerm
      \scriptscriptfont0=\sixrm
    \textfont1=\twelvei           \scriptfont1=\ninei
      \scriptscriptfont1=\sixi
    \textfont2=\twelvesy           \scriptfont2=\ninesy
      \scriptscriptfont2=\sixsy
    \textfont3=\twelveex          \scriptfont3=\ninex
      \scriptscriptfont3=\sixex
    \textfont\itfam=\twelveit    %\scriptfont\itfam=\nineit
    \textfont\slfam=\twelvesl    %\scriptfont\slfam=\ninesl
      \scriptscriptfont\slfam=\sixsl
    \textfont\bffam=\twelvebf     \scriptfont\bffam=\ninebf
      \scriptscriptfont\bffam=\sixbf
    \textfont\ttfam=\twelvett
    \textfont\cpfam=\twelvecp
    \textfont\ssfam=\twelvess     \scriptfont\ssfam=\niness
      \scriptscriptfont\ssfam=\sixss
    \ifamsfonts
       \textfont\msafam=\twelvemsa     \scriptfont\msafam=\ninemsa
         \scriptscriptfont\msafam=\sixmsa
       \textfont\msbfam=\twelvemsb     \scriptfont\msbfam=\ninemsb
         \scriptscriptfont\msbfam=\sixmsb
       \textfont\eufam=\twelveeu     \scriptfont\eufam=\nineeu
         \scriptscriptfont\eufam=\sixeu \fi
    \samef@nt
    \b@gheight=12pt
    \setbox\strutbox=\hbox{\vrule height 0.85\b@gheight
                                depth 0.35\b@gheight width\z@ }}
\twelvepoint
%
%%%%%%%%%%%%%%%%%
%%%Basic skips%%%
%%%%%%%%%%%%%%%%%
%
\baselineskip = 15pt plus 0.2pt minus 0.1pt %was 20pt ...
\lineskip = 1.5pt plus 0.1pt minus 0.1pt
\lineskiplimit = 1.5pt
\parskip = 6pt plus 2pt minus 1pt
\interlinepenalty=50
\interfootnotelinepenalty=5000
\predisplaypenalty=9000
\postdisplaypenalty=500
\hfuzz=1pt
\vfuzz=0.2pt
\dimen\footins=24 truecm % 8 truein in SB
\ifafour
 \hsize=16cm \vsize=22cm
\else
 \hsize=6.5in \vsize=9in
\fi
%
%%%%%%%%%%%%%%%
%%%Footnotes%%%
%%%%%%%%%%%%%%%
%
\skip\footins=\medskipamount
\newcount\fnotenumber
\def\clearfnotenumber{\fnotenumber=0} \clearfnotenumber
\def\fnote{\global\advance\fnotenumber by1 \generatefootsymbol
 \footnote{$^{\footsymbol}$}}
\def\fd@f#1 {\xdef\footsymbol{\mathchar"#1 }}
\def\generatefootsymbol{\iffrontpage\ifcase\fnotenumber
\or \fd@f 279 \or \fd@f 27A \or \fd@f 278 \or \fd@f 27B
\else  \fd@f 13F \fi
\else\xdef\footsymbol{\the\fnotenumber}\fi}
%
%%%%%%%%%%%%%%%%%%%%%%%%%%%%%
%%%Sections and Appendices%%%
%%%%%%%%%%%%%%%%%%%%%%%%%%%%%
%
\newcount\secnumber \newcount\appnumber
\def\clearappnumber{\appnumber=64} \def\clearsecnumber{\secnumber=0}
\clearsecnumber \clearappnumber
\newif\ifs@c % this is true if within a section as opposed to an appendix
\newif\ifs@cd % this is true if the article is being section'd
\s@cdtrue % this is the default
\def\unsectioned{\s@cdfalse\let\section=\subsection}
\newskip\sectionskip         \sectionskip=\medskipamount
\newskip\headskip            \headskip=8pt plus 3pt minus 3pt
\newdimen\sectionminspace    \sectionminspace=10pc
\def\Titlestyle#1{\par\begingroup \interlinepenalty=9999
     \leftskip=0.02\hsize plus 0.23\hsize minus 0.02\hsize
     \rightskip=\leftskip \parfillskip=0pt
     \advance\baselineskip by 0.5\baselineskip%this is a test...
     \hyphenpenalty=9000 \exhyphenpenalty=9000
     \tolerance=9999 \pretolerance=9000
     \spaceskip=0.333em \xspaceskip=0.5em
     \fourteenpoint
  \noindent #1\par\endgroup }
\def\titlestyle#1{\par\begingroup \interlinepenalty=9999
     \leftskip=0.02\hsize plus 0.23\hsize minus 0.02\hsize
     \rightskip=\leftskip \parfillskip=0pt
     \hyphenpenalty=9000 \exhyphenpenalty=9000
     \tolerance=9999 \pretolerance=9000
     \spaceskip=0.333em \xspaceskip=0.5em
     \fourteenpoint
   \noindent #1\par\endgroup }
\def\spacecheck#1{\dimen@=\pagegoal\advance\dimen@ by -\pagetotal
   \ifdim\dimen@<#1 \ifdim\dimen@>0pt \vfil\break \fi\fi}
\def\section#1{\cleareqnumber \s@ctrue \global\advance\secnumber by1
   \par \ifnum\the\lastpenalty=30000\else
   \penalty-200\vskip\sectionskip \spacecheck\sectionminspace\fi
   \noindent {\caps\enspace\S\the\secnumber\quad #1}\par
   \nobreak\vskip\headskip \penalty 30000 }
\def\undertext#1{\vtop{\hbox{#1}\kern 1pt \hrule}}
\def\subsection#1{\par
   \ifnum\the\lastpenalty=30000\else \penalty-100\smallskip
   \spacecheck\sectionminspace\fi
   \noindent\undertext{#1}\enspace \vadjust{\penalty5000}}

\def\appendix#1{\cleareqnumber \s@cfalse \global\advance\appnumber by1
   \par \ifnum\the\lastpenalty=30000\else
   \penalty-200\vskip\sectionskip \spacecheck\sectionminspace\fi
   \noindent {\caps\enspace Appendix \char\the\appnumber\quad #1}\par
   \nobreak\vskip\headskip \penalty 30000 }
\def\ack{\par\penalty-100\medskip \spacecheck\sectionminspace
   \line{\fourteencp\hfil ACKNOWLEDGEMENTS\hfil}%
\nobreak\vskip\headskip }
\def\refs{\begingroup \par\penalty-100\medskip \spacecheck\sectionminspace
   \line{\fourteencp\hfil REFERENCES\hfil}%
\nobreak\vskip\headskip \frenchspacing }
\def\endrefs{\par\endgroup}
\def\NoteAdded{\noindent{\caps Note added:}\enspace}%--- Note added
%
%%%%%%%%%%%%%%%%%%%%%%%%%%%%%%%%%
%%%Running heads and footlines%%%
%%%%%%%%%%%%%%%%%%%%%%%%%%%%%%%%%
%
\newif\iffrontpage \frontpagefalse
\headline={\hfil}
\footline={\iffrontpage\hfil\else \hss\twelverm
-- \folio\ --\hss \fi }
%
%%%%%%%%%%%%%%%%
%%%Title page%%%
%%%%%%%%%%%%%%%%
%
\newskip\frontpageskip \frontpageskip=12pt plus .5fil minus 2pt
\def\titlepage{\global\frontpagetrue\hrule height\z@ \relax
               \pubblock\relax }
\def\endtitlepage{\vfil\break\clearfnotenumber\frontpagefalse}
\def\title#1{\vskip\frontpageskip\Titlestyle{\caps #1}\vskip3\headskip}
\def\author#1{\vskip.5\frontpageskip\titlestyle{\caps #1}\nobreak}
\def\and{\par\kern 5pt \centerline{\sl and}}

\def\address#1{\par\kern 5pt\titlestyle{\it #1}}
\def\andaddress{\par\kern 5pt \centerline{\sl and} \address}

\def\abstract#1{\par\dimen@=\prevdepth \hrule height\z@ \prevdepth=\dimen@
   \vskip\frontpageskip\spacecheck\sectionminspace
   \centerline{\fourteencp ABSTRACT}\vskip\headskip
   {\noindent #1}}

\def\email#1{\fnote{\tentt e-mail: #1\hfill}}
\def\newaddress#1{\fnote{\tenrm #1\hfill}}
%
%%%%%%%%%%%%%%%%%%%%
%%%some addresses%%%
%%%%%%%%%%%%%%%%%%%%
%

%
\def\Bonn{\address{%
   Physikalisches Institut der Universit\"at Bonn\break
  Nu{\ss}allee 12, D--53115 Bonn, GERMANY}}
%

%

%
%%%%%%%%%%%%%%%%
%%%References%%%
%%%%%%%%%%%%%%%%
%
\newcount\refnumber \def\clearrefnumber{\refnumber=0}  \clearrefnumber
\newwrite\R@fs                              %This opens a file .refs with
\immediate\openout\R@fs=\jobname.refs %the references in order of
                                            %appearance.
\def\closerefs{\immediate\closeout\R@fs} %close file so that TeX can read it
\def\refsout{\closerefs\refs
\unlockat
\input\jobname.refs
\lockat
\endrefs}
\def\refitem#1{\item{{\bf #1}}}%just bolds it so that \bf does not expand
\def\ifundefined#1{\expandafter\ifx\csname#1\endcsname\relax}
\def\[#1]{\ifundefined{#1R@FNO}%
\global\advance\refnumber by1%
\expandafter\xdef\csname#1R@FNO\endcsname{[\the\refnumber]}%
\immediate\write\R@fs{\noexpand\refitem{\csname#1R@FNO\endcsname}%
\noexpand\csname#1R@F\endcsname}\fi{\bf \csname#1R@FNO\endcsname}}
\def\refdef[#1]#2{\expandafter\gdef\csname#1R@F\endcsname{{#2}}}
%
%%%%%%%%%%%%%%%
%%%Equations%%%
%%%%%%%%%%%%%%%
%
\newcount\eqnumber \def\cleareqnumber{\eqnumber=0}
\newif\ifal@gn \al@gnfalse  % this is true if within an \eqalignno
\def\veqnalign#1{\al@gntrue \vbox{\eqalignno{#1}} \al@gnfalse}
\def\eqnalign#1{\al@gntrue \eqalignno{#1} \al@gnfalse}
\def\(#1){\relax%
\ifundefined{#1@Q}
 \global\advance\eqnumber by1
 \ifs@cd
  \ifs@c
   \expandafter\xdef\csname#1@Q\endcsname{{%
\noexpand\rm(\the\secnumber .\the\eqnumber)}}
  \else
   \expandafter\xdef\csname#1@Q\endcsname{{%
\noexpand\rm(\char\the\appnumber .\the\eqnumber)}}
  \fi
 \else
  \expandafter\xdef\csname#1@Q\endcsname{{\noexpand\rm(\the\eqnumber)}}
 \fi
 \ifal@gn
    & \csname#1@Q\endcsname
 \else
    \eqno \csname#1@Q\endcsname
 \fi
\else%
\csname#1@Q\endcsname\fi\global\let\@Q=\relax}
%
%%%%%%%%%%%%%%%%%
%%%Mathematica%%%
%%%%%%%%%%%%%%%%%
%
\newif\ifm@thstyle \m@thstylefalse
\def\mathstyle{\m@thstyletrue}
\def\proclaim#1#2\par{\smallbreak\begingroup%        small --> med???
\advance\baselineskip by -0.25\baselineskip%
\advance\belowdisplayskip by -0.35\belowdisplayskip%
\advance\abovedisplayskip by -0.35\abovedisplayskip%
    \noindent{\caps#1.\enspace}{#2}\par\endgroup%
\smallbreak}%--- defs, thms, ...                     small --> med???
\def\m@kem@th<#1>#2#3{%
\ifm@thstyle \global\advance\eqnumber by1
 \ifs@cd
  \ifs@c
   \expandafter\xdef\csname#1\endcsname{{%
\noexpand #2\ \the\secnumber .\the\eqnumber}}
  \else
   \expandafter\xdef\csname#1\endcsname{{%
\noexpand #2\ \char\the\appnumber .\the\eqnumber}}
  \fi
 \else
  \expandafter\xdef\csname#1\endcsname{{\noexpand #2\ \the\eqnumber}}
 \fi
 \proclaim{\csname#1\endcsname}{#3}
\else
 \proclaim{#2}{#3}
\fi}
\def\Thm<#1>#2{\m@kem@th<#1M@TH>{Theorem}{\sl#2}}%--- Theorem
\def\Prop<#1>#2{\m@kem@th<#1M@TH>{Proposition}{\sl#2}}%--- Proposition
\def\Def<#1>#2{\m@kem@th<#1M@TH>{Definition}{\rm#2}}%--- Definition
\def\Lem<#1>#2{\m@kem@th<#1M@TH>{Lemma}{\sl#2}}%--- Lemma
\def\Cor<#1>#2{\m@kem@th<#1M@TH>{Corollary}{\sl#2}}%--- Corollary
\def\Conj<#1>#2{\m@kem@th<#1M@TH>{Conjecture}{\sl#2}}%--- Conjecture
\def\Rmk<#1>#2{\m@kem@th<#1M@TH>{Remark}{\rm#2}}%--- Remark
\def\Exm<#1>#2{\m@kem@th<#1M@TH>{Example}{\rm#2}}%--- Example
\def\Qry<#1>#2{\m@kem@th<#1M@TH>{Query}{\it#2}}%--- Query
%
%--- Proof
%

%
\def\<#1>{\csname#1M@TH\endcsname}
%
%%%%%%%%%%%%%%%%%%%
%%%Abbreviations%%%
%%%%%%%%%%%%%%%%%%%
%
\def\ref#1{{\bf [#1]}}%--- [ref]
%--- et al.
%--- i.e.
%--- e.g.
%--- Cf.
%--- cf.
 %--- double left quote
%--- th as in fifth
\def\nl{\hfil\break}%--- new line
%
%%%%%%%%%%%%%%%%%
%%%Mathematics%%%
%%%%%%%%%%%%%%%%%
%
%--- def over =
%--- Halmos Q.E.D.

%--- implies
%--- is implied by
%--- if and only if
\def\lapprox{\hbox{\lower3pt\hbox{$\buildrel<\over\sim$}}}% approx lt
\def\gapprox{\hbox{\lower3pt\hbox{$\buildrel<\over\sim$}}}% approx gt
\def\quotient#1#2{#1/\lower0pt\hbox{${#2}$}}%--- factor objects
\def\fr#1/#2{\mathord{\hbox{${#1}\over{#2}$}}}
\ifamsfonts
 \mathchardef\empty="0\hexmsb3F %--- better empty set than \emptyset
 \mathchardef\lsemidir="2\hexmsb6E % semidirect |x
 \mathchardef\rsemidir="2\hexmsb6F % semidirect x|
\else
 \let\empty=\emptyset
 \def\lsemidir{\mathbin{\hbox{\hskip2pt\vrule height 5.7pt depth -.3pt
    width .25pt\hskip-2pt$\times$}}}
 \def\rsemidir{\mathbin{\hbox{$\times$\hskip-2pt\vrule height 5.7pt
    depth -.3pt width .25pt\hskip2pt}}}
\fi
%
%--- injective map
%--- surjective map
%--- bijective map
%--- mapping
%--- long mapping
%--- isom over -->
%--- just an abbrev.
%

%
 %--- commutative diagram macro
 %--- map in complex
%
 %--- reals
 %--- complex nos.
 %--- quaternions
 %--- integers
 %--- rationals
 %--- naturals
 %--- ground field
%
%--- Hom(omorphisms)
%--- tr(ace)
%--- Tr(ace)
%--- End(omorphisms)
%--- Mor(phisms)
%--- Aut(omorphisms)
%--- aut(omorphisms)
%--- supertrace
%--- superdeterminant
%--- kernel
%--- cokernel
%--- image
%
\def\underrightarrow#1{\vtop{\ialign{##\crcr
      $\hfil\displaystyle{#1}\hfil$\crcr
      \noalign{\kern-\p@\nointerlineskip}
      \rightarrowfill\crcr}}} %--- modification of \overrightarrow
\def\underleftarrow#1{\vtop{\ialign{##\crcr
      $\hfil\displaystyle{#1}\hfil$\crcr
      \noalign{\kern-\p@\nointerlineskip}
      \leftarrowfill\crcr}}}  %--- modification of \overleftarrow

\def\comm#1#2{\left[#1\, ,\,#2\right]}%--- [ , ]
%--- { , }
%--- [ , }
%
%--- Lie derivative
%--- vartnl derivative
%--- partial derivative
%--- full derivative
%
%%%%%%%%%%%%%%
%%%Journals%%%
%%%%%%%%%%%%%%
%

\def\NPB#1#2#3{{\sl Nucl. Phys.} {\bf B#1} (#2) #3}

\def\CMP#1#2#3{{\sl Comm. Math. Phys.} {\bf #1} (#2) #3}

\def\PLB#1#2#3{{\sl Phys. Lett.} {\bf #1B} (#2) #3}

\def\UMN#1#2#3{{\sl Usp. Mat. Nauk} {\bf #1} (#2) #3}
\def\RMS#1#2#3{{\sl Russian Math Surveys} {\bf #1} (#2) #3}

\def\IJMPA#1#2#3{{\sl Int. J. Mod. Phys.} {\bf A#1} (#2) #3}

\def\MPLA#1#2#3{{\sl Mod. Phys. Lett.} {\bf A#1} (#2) #3}

\lockat
%
%%%%%%%%%%%%%%%%%%%%%%%%%%%%%%%%%%%%%%%%%%%%%%%%%%%%%%%%%%%%%
%
%   These are the local macros for ITP-SB-89-NOGHOST
%

\def\W{\mathord{\ssf W}}
\def\WB{\mathord{\ssf WB}}
\def\ope[#1][#2]{{{#2}\over{\ifnum#1=1 {z-w} \else {(z-w)^{#1}}\fi}}}

\def\reg{\mathord{\hbox{reg.}}}

\def\Jgh{\mathord{J^{\rm gh}}}
\def\Jtot{\mathord{J^{\rm tot}}}
\def\gg{\mathord{\eu g}}
\def\gh{\mathord{\eu h}}
\def\gghat{\widehat{\gg}}
\def\ghhat{\widehat{\gh}}

\def\fr#1/#2{\mathord{\hbox{${#1}\over{#2}$}}}
\def\half{\fr1/2}
\let\d=\partial
\def\hepth/#1/{{\tt hep-th/#1}}
\def\using[#1]{&\hbox{by }#1}
\def\Jt{\mathord{\widetilde{J}}}
\def\Jh{\mathord{\widehat{J}}}
\def\comm[#1,#2]{\mathord{\left[#1,#2\right]}}
\refdef[KS]{D. Karabali and H. J. Schnitzer, \NPB{329}{1990}{649}.}
\refdef[Akman]{F. Akman, {\sl A characterization of the differential
in semi-infinite cohomology}, \hepth/9302141/.}
\refdef[BBSS]{F.A. Bais, P. Bouwknegt, K. Schoutens, and M. Surridge,
\NPB{304}{1988}{371}.}
\refdef[Fe]{B.L. Feigin, {\it The semi-infinite homology of Kac-Moody and
Virasoro Lie algebras}, \UMN{39}{1984}{195} (English Translation:
\RMS{39}{1984}{155}).}
\refdef[vNSS]{P. van Nieuwenhuizen, K. Schoutens, and A. Sevrin,
\CMP{124}{1989}{87}.}
\refdef[Horn]{K. Hornfeck, {\sl Explicit construction of the BRST
charge for $\W_4$}, \hepth/9306019/.}
\refdef[Zhu]{C.-J. Zhu, {\sl The BRST quantization of the nonlinear
$\WB_2$ and $\W_4$ algebras}, \hepth/9306026/.}
\refdef[QDS]{B. Feigin and E. Frenkel, \PLB{246}{1990}{75};\nl
J.M. Figueroa-O'Farrill, \NPB{343}{1990}{450};\nl
J. de Boer and T. Tjin, {\sl The relation between quantum
$\W$-algebras and Lie algebras}, \hepth/9302006/.}
\refdef[TFT]{E. Witten, \CMP{117}{1988}{353}, \CMP{118}{1988}{411},
\NPB{340}{1990}{281};\nl
T. Eguchi and S. Yang, \MPLA{4}{1990}{1693}.}
\refdef[GRS]{B. Gato-Rivera and A.M. Semikhatov, \PLB{293}{1992}{72}\nl
(\hepth/9207004/).}
\refdef[MV]{S. Mukhi and C. Vafa, {\sl Two-dimensional black hole as a
topological coset model of $c{=}1$ string theory}, \hepth/9301083/.}
\refdef[BLNW]{M. Bershadsky, W. Lerche, D. Nemeshansky, and N. Warner,
{\sl Extended $N{=}2$ superconformal structure of gravity and
$\W$-gravity coupled to matter}, \hepth/9211040/.}
\refdef[wiiineqii]{C. Pope, private communication;\nl
J.M. Figueroa-O'Farrill and K. Thielemans, unpublished.}
\refdef[Romans]{L. Romans, \NPB{352}{1991}{829}.}
\refdef[KaSu]{Y. Kazama and H. Suzuki, \PLB{216}{1989}{112},
\NPB{321}{1989}{232}.}
\refdef[Wi]{E. Witten, \NPB{371}{1992}{191}.}
\refdef[DT]{A. Deckmyn and W. Troost, \NPB{370}{1992}{231}.}
\refdef[FiMo]{J.M. Figueroa-O'Farrill and N. Mohammedi, work in
progress.}
\refdef[FMO]{J.M. Figueroa-O'Farrill, N. Mohammedi, and N. Obers, work
in progress.}
\refdef[GoSch]{P. Goddard and A. Schwimmer, \PLB{214}{1988}{209}.}
\refdef[DeTh]{A. Deckmyn and K. Thielemans, {\sl Factoring out free
fields},\nl \hepth/9306129/.}
\refdef[GHKO]{A. Giveon, M.B. Halpern, E. Kiritsis, and N. Obers,
\IJMPA{7}{1992}{947}.}
\refdef[HuSp]{C.M. Hull and B. Spence, \PLB{241}{1990}{357}.}
\refdef[getzler]{E. Getzler, {\sl Manin triples and $N{=}2$
superconformal field theory}, MIT Math Preprint.}
\refdef[leuven]{A. Sevrin, Ph. Spindel, W. Troost and A. Van Proeyen,
\NPB{308}{1988}{662}, \NPB{311}{1988/89}{465}.}
\overfullrule=0pt
\def\pubblock{ \line{\hfil\rm BONN--HE--93--20}
               \line{\hfil\tt hep-th/9306164}
               \line{\hfil\rm June 1993}}
\titlepage
\title{Affine Algebras, $N{=}2$ Superconformal Algebras, and
Gauged WZNW Models}
\author{Jos\'e~M.~Figueroa-O'Farrill
\email{jmf@avzw01.physik.uni-bonn.de}
\newaddress{Address after September 1, 1993: Physics Department, Queen
Mary and\nl Westfield College, London, UK.}}
\Bonn
\abstract{We find a canonical $N{=}2$ superconformal algebra (SCA)
in the BRST complex associated to any affine Lie algebra $\ghhat$ with
$\gh$ semisimple.  In contrast with the similar known results for the
Virasoro, $N{=}1$ supervirasoro, and $\W_3$ algebras, this SCA does
not depend on the particular ``matter'' representation chosen.
Therefore it follows that every gauged WZNW model with data
$(\gg\supset\gh, k)$ has an $N{=}2$ SCA with central charge
$c=3\dim\gh$ independent of the level $k$.  In particular, this
associates to every embedding $sl(2) \subset \gg$ a one-parameter
family of $c{=}9$ $N{=}2$ supervirasoro algebras.  As a by-product of
the construction, one can deduce a new set of ``master equations'' for
generalized $N{=}2$ supervirasoro constructions which is simpler than
the one considered thus far.}
\endtitlepage

\section{Introduction}

For some time now it has been realized that to any conformal field
theory with $N{=}2$ superconformal symmetry one can associate a
cohomology theory which is intimately related to the topological
conformal field theory one obtains after twisting\[TFT].  The converse
statement---namely, that in any ``reasonable'' cohomology theory one
can find an underlying $N{=}2$ SCA is not so firmly established, based
mostly on indirect arguments and special examples taken from string
theory.  These examples suffer from one major drawback: the $N{=}2$
SCA is not at all evident and finding it requires a tedious and
unenlightening albeit systematic procedure.  Moreover the $N{=}2$
depends on the ``matter'' representation and hence the construction is
not canonical.  The purpose of this paper is to describe a {\sl
natural} $N{=}2$ SCA present in the BRST complex associated to any
(untwisted) affine algebra $\ghhat$ with $\gh$ semisimple.  In
particular, this means that every gauged WZNW model with semisimple
gauged subgroup has an $N{=}2$ SCA whose central charge, as we will
see, is fixed to $c=3\dim\gh$.

To place this result in context, let us first consider the $N{=}2$
supervirasoro algebra.  It is generated by fields $J$ $,G^\pm$ and
$T$, where $T$ generates a Virasoro algebra relative to which $J$ and
$G^\pm$ are primary fields of weights $1$ and $\fr3/2$ respectively.
The remaining OPEs are given by
$$
\eqnalign{
J(z)J(w) &= \ope[2][c/3] + \reg~, \(JJope)\cr
J(z)G^\pm(w) &= \pm \ope[1][G^\pm(w)] + \reg~, \(JGope)\cr
G^\pm(z) G^\pm(w) &= \reg~, \(GpmGpmope)\cr
\noalign{\hbox{and}}
G^\pm(z) G^\mp(w) &= \ope[3][2c/3] \pm \ope[2][2J(w)] + \ope[1][2T(w)
\pm J(w)] + \reg~. \(GpmGmpope)\cr}
$$
In particular, \(GpmGpmope) implies that the charge $Q$ corresponding
to $G^+(z)$ ($Q=G^+_{-1/2}$ in the Neveu-Schwarz sector and
$Q=G^+_{-1}$ in the Ramond) squares to zero.  The resulting cohomology
defines the chiral ring of the theory.

Relative to the twisted energy-momentum tensor $T_{\rm twisted} \equiv
T + \fr1/2 \d J$, the spectrum of the fields changes, since the
conformal weight receives contributions from the charge.  Hence
positively (resp. negatively) charged fields decrease (resp. increase)
their conformal weight by half of its charge, while neutral fields
weigh the same.  This means that $G^+$ is now a primary field of
weight $1$, whereas $G^-$ has weight $2$.  Moreover, $T_{\rm
twisted}(z) = \half \comm[Q,G^-(z)]$.  This is very reminiscent of
critical string theory, where the roles of $J$, $G^+$, $G^-$, and
$T_{\rm twisted}$ are played by the ghost number current, the BRST
current, the antighost, and the total energy-momentum tensor,
respectively.  However the analogy seems to break down since, for
example, the BRST current does not have a regular OPE with itself.
The crucial observation is that the BRST current is defined up to a
total derivative and that in some cases (for example, if there is a
free boson in the theory) one can modify the BRST current in such a
way that the singularities in the OPE with itself cancel.  This also
induces a modification in the ghost number current.

The resulting $N{=}2$ algebra was shown explicitly for the first time
in the literature in \[GRS] (see also \[MV]) for the $c{=}1$
noncritical string.  In \[BLNW] this observation was generalized and
argued to be a generic property of string theories: be it the
``humble'' string, the superstring, or the $\W$-string.  Explicit
constructions of the $N{=}2$ algebra depend on the model under
consideration and were given in \[BLNW] for the bosonic string
(critical and noncritical), the NSR string (critical and
noncritical)---where one actually has an $N{=}3$ SCA---and for the
noncritical $\W_3$-string, where one finds an extension of $N{=}2$ by
an $N{=}2$ primary of weight $2$ and charge $0$.  For the critical
$\W_3$ string with matter representation given by the Romans
realization of $\W_3$ in terms of a free boson and an underlying
Virasoro algebra \[Romans] one can also construct such an $N{=}2$
extended algebra \[wiiineqii] whose central charge is fixed to
either of the two values $c=-18$ or $c=-\fr15/2$.  It is expected that
this continues to be the case for other $\W$-string theories built on
other $\W$-algebras; but proving it by the current means requires
knowing (at least the existence of) the BRST charge for the relevant
$\W$-algebra, for which no general construction is known and hence
must be done case by case.  The computational complexity soon becomes
forbidding (even to a computer) and so far the only other algebras
whose BRST charge is known are: $\W_4$ \[Horn] \[Zhu], $\WB_2$ \[Zhu]
and some special quadratically nonlinear algebras \[vNSS].  One can
certainly envision a few more algebras to be reached in the near
future, but this is far from a general existence proof.

Since $\W$-algebras are generically defined via the quantum
Drinfel'd-Sokolov reduction \[QDS], it is widely believed that the BRST
charge and the underlying $N{=}2$ structure should also come induced
from the analogous structures in the affine Lie algebra from which one
reduces.  The BRST charge for an (untwisted) affine algebra $\ghhat$
is as old as semi-infinite cohomology itself.  It corresponds to the
differential in the subcomplex of the Feigin standard complex
computing the semi-infinite cohomology of $\ghhat$ relative to the
center, and its definition goes back to Feigin \[Fe]; although it is
not inconceivable that it could be found already in the earlier physics
literature.

In this paper we construct the underlying $N{=}2$ structure for the
case of $\gh$ semisimple.  No canonical construction seems to exist
for $\gh$ abelian; although depending on the ``matter''
representation, an $N{=}2$ SCA sometimes exist.  Since our construction
is otherwise completely general, it means that this $N{=}2$ structure
is present in any gauged WZNW model with semisimple gauged subgroup.

The plan of this paper is as follows.  In section 2 we will review the
construction of the BRST charge for an affine Lie algebra $\ghhat$
with $\gh$ semisimple and define the other generators of the algebra.
In a nutshell, the roles of $J$, $G^+$ and $G^-$ are played by the
ghost number, BRST and antiBRST currents, respectively.  In section 3
we prove that they correspond indeed to an $N{=}2$ SCA.  In doing so
we have found what we believe is the minimal data necessary to
guarantee the existence of an $N{=}2$ supervirasoro algebra.  The
results of this section could thus be useful in the context of
generalized $N{=}2$ supervirasoro constructions.  In section 4 we
apply this to the gauged WZNW model with data $(\gg\supset\gh,k)$.
The construction yields for fixed $(\gg\supset\gh)$ a family of
$N{=}2$ SCA's with central charge $c=3\dim\gh$ indexed by the level
$k$.  Section 5 summarizes the results of the paper.

\section{The BRST Complex of an Affine Lie Algebra}

In this section we describe the BRST complex for an affine Lie
algebra.  Let us consider a semisimple Lie algebra $\gh$ with a fixed
invariant nondegenerate bilinear form $\gamma$.  Let us (reluctantly)
introduce a basis $\{X_i\}$ for $\gh$ relative to which the structure
constants will be denoted by $f_{ij}^k$ and such that the bilinear
form $\gamma$ has entries $\gamma_{ij}$.  We let $\gamma^{ij}$ denote
the inverse to $\gamma_{ij}$, and $c_{\gh}$ denote the eigenvalue of
the Casimir element $\gamma^{ij} X_iX_j$ on the adjoint
representation; in other words,
$$
f_{ik}^l f_{jl}^k = c_{\gh} \gamma_{ij}~.                     \()
$$
We let $\ghhat$ denote the (untwisted) affine algebra associated to
$\gh$.  It is spanned by currents $\{J_i(z)\}$ obeying the following
OPE:
$$
J_i(z) J_j(w) = \ope[2][-c_{\gh} \gamma_{ij} ] + \ope[1][f_{ij}^k
J_k(w)] + \reg~.                                         \(affineh)
$$
We have fixed the level to $-c_{\gh}$ for reasons that are standard
and which will become apparent presently.  The other ingredients are
the (fermionic) ghost fields $(b_i,c^i)$ with the standard OPE:
$$
b_i(z) c^j(w) = \ope[1][\delta_i^j] + \reg~.            \(ghostope)
$$
One can build $\ghhat$ currents out of ghost bilinears $\Jgh_i(z)
\equiv f_{ij}^k b_k c^j(z)$.\fnote{Products of fields at the same
point are assumed to be normal-ordered according to the
point-spittling convention and, moreover, normal-ordering associates
to the left (see, for example, \[BBSS]).}  The ghost currents satisfy
the OPE
$$
\Jgh_i(z) \Jgh_j(w) = \ope[2][c_{\gh} \gamma_{ij} ] + \ope[1][f_{ij}^k
\Jgh_k(w)] + \reg~.                                 \(ghostcurrents)
$$
Notice that the total currents $\Jtot_i = J_i + \Jgh_i$ represent
$\ghhat$ at level 0.  This fact is equivalent (see, for example,
\[Akman]) to the ``nilpotency'' of the BRST charge $Q$, which is the
charge associated to the BRST current
$$
G^+ (z) \equiv {2i\over\sqrt{c_{\gh}}} \left( c^iJ_i(z) + \half c^i
\Jgh_i(z)\right)~,                                        \(Gplus)
$$
where we have chosen a convenient normalization.  The equation $Q^2=0$
is equivalent to the fact that the first order pole in the OPE $G^+(z)
G^+(w)$ is a total derivative.  Here, however, we find the stronger
result that the OPE $G^+(z) G^+(w)$ is regular.  This is to be
contrasted with the BRST current in ($\W$-)string theory in which one
finds singularities in its OPE with itself.  Depending on the matter
content of the string theory into consideration, these singularities
can be cancelled by modifying the BRST current with total derivative
pieces.  But this makes the construction of the associated $N{=}2$ SCA
rather noncanonical.

Let us now introduce the ghost number current
$$
J(z) \equiv c^ib_i(z)~.                            \(ghostnumber)
$$
A quick calculation shows that it obeys the following OPE
$$
J(z) G^+(w) = \ope[1][G^+(w)] + \reg~.             \(JGplus)
$$
We now introduce an automorphism $\pi$ of our operator product
algebra:
$$\pi(b_i) = \gamma_{ij}c^j \qquad \pi(c^i) = \gamma^{ij}b_j \qquad
\pi(J_i) = J_i~.                                    \(auto)
$$
It is an involutive automorphism that fixes the ghost currents
$\pi(\Jgh_i) = \Jgh_i$.  Since it is essentially ghost conjugation, it
obeys $\pi(J) = -J$.  Let us define $G^- \equiv \pi(G^+)$; or,
explicitly,
$$
G^- (z) \equiv {2i\over\sqrt{c_{\gh}}} \gamma^{ij}\left( b_iJ_j(z) +
\half b_i \Jgh_j(z)\right)~.                        \(Gminus)
$$
Because $\pi$ is an automorphism, we immediately have that
$$
J(z) G^-(w) = \ope[1][-G^-(w)] + \reg~,             \(JGminus)
$$
and also
$$
G^-(z) G^-(w) = \reg~.                              \()
$$
A calculation now shows that
$$
G^+(z) G^-(w) = \ope[3][2c/3] + \ope[2][2J(w)] + \ope[1][2T(w) + \d
J(w)] + \reg~,                                       \()
$$
where $c = 3\dim\gh$ and $T$ is found to be
$$
T = -{1\over c_{\gh}} \gamma^{ij}\left(2 J_iJ_j + J_i\Jgh_j +
\Jgh_iJ_j + \Jgh_i\Jgh_j\right) + \half \left(\d b_i c^i - b_i \d
c^i\right)~.                                        \(EMtensor)
$$
We will see in the next section that these results together with the
associativity of the underlying operator algebra, already imply that
$(J,G^\pm,T)$ generate an $N{=}2$ SCA with central charge $c =
3\dim\gh$.

The form of $T$ may seem at first strange, but let us notice some
things.  The last term is nothing but the energy-momentum tensor for
free fermions $b_i$ and $c^i$.  In terms of the total currents
$\Jtot_i$, we can rewrite the rest of $T$ as follows
$$
-{1\over c_{\gh}}\gamma^{ij} (\Jtot_i \Jtot_j + J_i J_j)\(EMweird)
$$
The first term is---up to a sign---the Sugawara tensor for the
total currents which represent $\ghhat$ at level zero; whereas the
second term is proportional to what would be the Sugawara tensor for
the $J_i$, {\sl only} that these currents are at the critical level,
where the Sugawara construction breaks down.  One should stress that
whereas the last term of \(EMtensor) generates a Virasoro algebra on
its own, \(EMweird) does not.  It is nevertheless interesting to
remark that energy-momentum tensors of this form have appeared in the
literature \[DT] as generalized Virasoro constructions based on affine
Lie superalgebras.  In fact, what appears in \[DT] is a one-parameter
family of such tensors.  Perhaps ``affine Virasoro space'' at the
critical level deserves a closer study.

\section{When do we have an $N{=}2$ SCA?}

In this section we prove a result, which may be of
independent interest in the context of generalized $N{=}2$
constructions.  It concerns the minimal data necessary to guarantee
that we have an $N{=}2$ SCA.  We tacitly assume the associativity of
the operator product expansion.  Suppose we have fields $G^\pm$
satisfying \(GpmGpmope) and
$$
G^+(z) G^-(w) = \ope[3][2c/3] + \ope[2][2J(w)] + \ope[1][2T(w) + \d
J(w)] + \reg~,                                          \(GpGm)
$$
which {\it defines} $c$, $J$, and $T$.  Then if, in addition, \(JGope)
is also obeyed, then $(J,G^\pm,T)$ satisfies an $N{=}2$ SCA with
central charge $c$.  In particular, $T$ is guaranteed to satisfy a
Virasoro algebra with central charge $c$.

The proof exploits the associativity of the operator product expansion
to compute the remaining OPEs and verify that they are indeed the ones
of an $N{=}2$ SCA.  It will prove convenient to introduce the
following notation.  If $A$ and $B$ are any two local fields, their
OPE defines a family of bilinears $[~]_\ell$ by
$$
A(z)B(w) = \sum_{\ell} \ope[\ell][{[AB]_\ell(w)}]~.          \(OPE)
$$
The associativity of the operator product expansion implies certain
well-known identities between nested bilinears.  In particular, a very
useful one is
$$
[A[BC]_n]_m = (-)^{|A||B|}[B[AC]_m]_n + \sum_{\ell=0}^{m-1} {m-1
\choose \ell} [[AB]_{m-\ell}C]_{n+\ell}~,        \(CauchyJacobi)
$$
which is valid for all $A,B,C$ and all positive $n,m$.  There are
similar identities for all $m$ and $n$ but we shall not need them.
Putting $m=1$ in \(CauchyJacobi), we find that for all $A$ the
operation $B \mapsto [AB]_1$ is a superderivation over the operator
product.

We begin by computing $J(z)J(w)$.  From \(GpGm) we see that $J =
\fr1/2 [G^+G^-]_2$.  Therefore,
$$
\eqnalign{[JJ]_n
&= \fr1/2 [J[G^+G^-]_2]_n \cr
&= \fr1/2 [G^+[JG^-]_n]_2 + \fr1/2 \sum_{\ell=0}^{n-1} {n-1 \choose
\ell} [[AB]_{n-\ell}C]_{2+\ell} \using[\(CauchyJacobi)]\cr
&= - \fr1/2 \delta_{n1} [G^+ G^-]_2 + \fr1/2 [G^+ G^-]_{n+1}~,\cr}
$$
from which we read off
$$
J(z)J(w) = \ope[2][c/3] + \reg ~,                        \(JJ)
$$
agreeing with \(JJope).

We now compute $T(z)J(w)$.  From \(GpGm) we find that $T = \fr1/2
[G^+G^-]_1 - \fr1/2 \d J$, whence
$$
\eqnalign{[JT]_n
&= \fr1/2 [J[G^+G^-]_1]_n - \fr1/2 [J \d J]_n \cr
&= \fr1/2 [G^+[J G^-]_n]_1 - \fr1/2 [[G^+ J]_1G^-]_n - c/3 \delta_{n3}
                                 \using[\(CauchyJacobi)] \cr
&= -\fr1/2 \delta_{n1} [G^+G^-]_1 + \fr1/2 [G^+ G^-]_n - c/3
                                                   \delta_{n3}~, \cr}
$$
from where we can read off
$$
T(z)J(w) = \ope[2][J(w)] + \ope[1][\d J(w)] + \reg ~.       \(TJ)
$$

Let us now compute $T(z)G^\pm(w)$.  We shall do $T(z)G^+(w)$ first.
The result for $T(z)G^-(w)$ will then follow after applying the
involutive automorphism $J\mapsto -J$, $G^\pm \mapsto G^\mp$, $T\mapsto
T$ of the abstract algebra defined by \(JGope), \(GpmGpmope), and \(GpGm).
Substituting for $T$ as before, we obtain
$$
\eqnalign{[G^+T]_n
&= \fr1/2 [G^+[G^+G^-]_1]_n - \fr1/2 [G^+ \d J]_n \cr
&= -\fr1/2 [G^+[G^+G^-]_n]_1 - \fr1/2 [G^+ \d J]_n
             \using[\(CauchyJacobi),\(GpmGpmope)]\cr}
$$
{}From this we read that $[G^+T]_2 = \fr3/2 G^+$ and $[G^+T]_1 = \fr1/2
\d G^+$ are the only nonzero terms in the singular part of the OPE.
Therefore, and after applying the automorphism,
$$
T(z)G^\pm(w) = \ope[2][\fr3/2 G^\pm(w)] + \ope[1][\d G^\pm(w)] +
\reg~.                                                    \(TGpm)
$$

Finally we compute $T(z)T(w)$.  Substituting for $T$ once, we again
find
$$
\eqnalign{[TT]_n
&= \fr1/2 [T[G^+G^-]_1]_n - \fr1/2 [T\d J]_n\cr
&= \fr1/2 [G^+[TG^-]_n]_1 - \fr1/2 [[G^+T]_1G^-]_n - \fr1/2 [T\d J]_n
                                              \using[\(CauchyJacobi)]\cr
&= \fr1/2 [G^+[TG^-]_n]_1 - \fr1/4 [\d G^+ G^-]_n - \fr1/2[T\d J]_n~,
                                                                   \cr
&= \fr3/4 \delta_{n2} [G^+ G^-]_1 + \fr1/2 \delta_{n1} [G^+\d G^-]_1
  - \fr1/4 [\d G^+ G^-]_n \cr
&\qquad - \delta_{n3} J - \delta_{n2} \d J - \fr1/2 \delta_{n1}
\d^2J~,\cr}
$$
which expands to the expected
$$
T(z)T(w) = \ope[4][c/2] + \ope[2][2T(w)] + \ope[1][\d T(w)] + \reg ~.
                                                             \(TT)
$$

We now briefly comment on how this result can be helpful in the search
for generalized $N{=}2$ supervirasoro constructions.  Generalized
$N{=}2$ supervirasoro constructions starting from Lie algebraic data
have not been subject to the same intensive study as its $N{=}0$ and
$N{=}1$ counterparts.  The only reference seems to be an appendix of
\[GHKO] where a set of ``master equations'' are written down which
are sufficient for the construction of an $N{=}2$ supervirasoro
algebra out of a Ka{\v c}--Todorov algebra with data $(\gg, k)$.  As a
result of trying to construct an $N{=}2$ algebra with $N{=}1$
ingredients, one finds that no natural Sugawara-type solution exists.
Instead one finds only the Kazama-Suzuki construction.  This is not
surprising and perhaps one should attempt the construction from an
$N{=}2$ extension of an affine Lie algebra, of the type considered by
Hull and Spence \[HuSp].  Work on this is in progress \[FMO].
Nevertheless, in order to make contact with the existing literature,
we shall present a simplified set of $N{=}2$ master equations on a
Ka{\v c}--Todorov algebra.  There is no reason to expect that the
$N{=}2$ SCA under construction should preserve the $N{=}1$ present in
the Ka{\v c}--Todorov algebra, so we shall not insist in working in
$N{=}1$ superfields, but in components.

The Ka{\v c}--Todorov algebra associated to the data $(\gg,k)$ is the
extension of $\gghat$ at level $k$ (represented by the currents
$J_i(z)$) by $\dim\gg$ free fermions $(\psi_i(z)$) transforming in the
adjoint representation of $\gg$.  The algebra is defined by the
following OPE's
$$
\eqnalign{
J_i(z) J_j(w) &= \ope[2][k g_{ij}] + \ope[1][f_{ij}^k J_k(w)] +
\reg~,\cr
J_i(z) \psi_j(w) &= \ope[1][f_{ij}^k \psi_k(w)] + \reg~,\cr
\psi_i(z) \psi_j(w) &= \ope[1][k g_{ij}] + \reg~.\cr}
$$
According to general decoupling arguments (see \[DeTh] generalizing
the classical result of Goddard-Schwimmer \[GoSch]), one can decouple
the fermions by shifting the currents $J_i \mapsto J_i - \fr1/{2k}
f_{ij}^k g^{jl} \psi_k \psi_l$.  Whereas decoupled fermions may
simplify some of the calculations, notice that the shift is illegal
for $k=0$.  Since we expect that something should happen there, we
will leave the fermions coupled.

According to the results in this section, one need only write the most
general $G^\pm$:
$$
G^\pm = A_\pm^{ij} \psi_i J_j + B_\pm^{ijk}\psi_i\psi_j\psi_k +
        C_\pm^i \d\psi_i~,\(gg)
$$
where $B_\pm^{ijk}$ is totally antisymmetric and $A_\pm^{ij}$ is
general.   The first equations come from imposing \(GpmGpmope).
Computing $G^+(z)G^-(w)$ one finds that the central charge is given
simply by $c = \fr3/2 [G^+G^-]_3$, whereas $J$ is given by $J =
\fr1/2[G^+G^-]_2$.  The final set of equations comes from demanding
\(JGope).  From the results of this section this is sufficient to
guarantee an $N{=}2$ supervirasoro algebra.  This not being the main
point of the paper, we will refrain from writing the equations down
explicitly.  The interested reader should find it a good exercise to
reproduce them.

\section{The $N{=}2$ SCA in the Gauged WZW Model}

Gauged WZNW models provide a very large class of examples to
which the construction in Section 2 applies.  We briefly review the
algebraic setup of the WZNW model and we refer the reader to \[KS] for
more details.  Algebraically, the gauged WZW model consists of three
independent theories coupled by constraints which are treated \`a la
BRST.  Consider a pair of finite-dimensional semisimple  Lie algebras
$\gh \subset \gg$.  For ease of exposition we treat the case where the
index of embedding is 1.  Clearly the construction goes through in the
general case.  Let us consider their affine counterparts $\ghhat
\subset \gghat$.  We again choose a basis $\{X_i\}$ for $\gh$ which we
complete to a basis $\{X_a\}$ for $\gg$.  We consider an invariant
non-degenerate symmetric bilinear form $\gamma$ on $\gg$.  Relative to
the basis $\{X_a\}$, $\gamma$ has entries $\gamma_{ab}$ and its
restriction to $\gh$ has entries $\gamma_{ij}$.  Consider then the
algebra of currents of $\gg$ at level $k$:
$$
\Jh_a(z) \Jh_b(w) = \ope[2][k \gamma_{ab} ] + \ope[1][f_{ab}^c
\Jh_c(w)] + \reg~.                                         \(affineg)
$$
This is our first theory: it corresponds to a (nongauged) WZNW model
with algebra $\gg$ at level $k$.  The second ingredient is another
WZNW model with algebra $\gh$ at level $-k-c_{\gh}$ and described by
currents
$$
\Jt_i(z) \Jt_j(w) = \ope[2][(-k-c_{\gh}) \gamma_{ij} ] +
\ope[1][f_{ij}^k \Jt_k(w)] + \reg~.                  \(affineht)
$$
The third and final theory are the ghosts $(b_i,c^i)$ obeying
\(ghostope).  The constraints are then simply the currents $J_i =
\Jh_i + \Jt_i$ which precisely satisfy \(affineh).  The data
$(J_i,b_i,c^i)$ is precisely the data that we need in order for the
construction in Section 2 to work.  Thus we conclude that in every
gauged WZNW with data $(\gg\supset\gh, k)$ with $\gh$ semisimple,
there is a canonical $N{=}2$ superconformal algebra with $c=3\dim\gh$.
Notice that this is independent of the level $k$, hence in effect it
gives a one-parameter family of $N{=}2$ SCA's associated to every
$\gg\supset\gh$.  In the case of $\gh$ some real form of $sl(2)$, we
find an infinite series of one-parameter families of $c{=}9$ theories,
each family associated to an embedding $sl(2)\subset \gg$.  This same
data seems to comprise the ``moduli space'' of $\W$-algebras.  Both
this fact and the fact that $c{=}9$ is the phenomenologically
interesting central charge in string theory are very intriguing and
certainly deserve further study.

Finally, one should hasten to add that the $N{=}2$ SCA does not
descend to the physical states; that is, it is not a symmetry of the
coset model to which the gauged WZNW is equivalent.  In fact, one of
the generators is the BRST current which is BRST exact: $G^+ =
-\comm[Q,J]$.  For the same reason, neither does the twisted
energy-momentum tensor $T_{\rm twisted} = T + \half \d J =
\comm[Q,G^-]$ descend. This is not surprising since not every coset
theory has $N{=}2$ superconformal symmetry: it seems, in fact, that in
the context of supersymmetric gauged WZNW models at least, it requires
$G/H$ to admit a  K\"ahler structure \[KaSu] \[Wi].  Nevertheless, it
may be possible, via the identification between the chiral ring of
the $N{=}2$ SCA and the physical states in the gauged WZNW model
(equivalently, the coset model), that results in one area can help
understand the other.

\section{Conclusion and Summary}

To recapitulate, we have shown the existence of a {\sl natural}
$N{=}2$ supervirasoro algebra in the BRST complex associated to any
(untwisted) affine algebra $\ghhat$ with semisimple $\gh$.  Unlike the
previous cases of $N{=}2$ SCA's found in BRST complexes, this one is
canonical and does not depend on the ``matter'' representation of the
current algebra.  Therefore this construction applies to any gauged
WZNW model with data $(\gh \subset \gg,k)$ with $\gh$ semisimple.  The
central charge of the $N{=}2$ SCA is equal to $3\dim\gh$ which is
independent of the level $k$.  For $\gh = sl(2)$, this construction
associates a one-parameter ($k$) family of $c{=}9$ $N{=}2$ SCA's to
any embedding $sl(2)\in\gg$---a highly intriguing fact since it seems
to connect the ``moduli space'' of $\W$-algebras to the
phenomenologically interesting string theories with $c{=}9$ $N{=}2$
superconformal symmetry.

As a more technical by-product of this construction, we have given
what we think is the minimal data required to guarantee an $N{=}2$
supervirasoro algebra.  This should simplify the search of generalized
$N{=}2$ supervirasoro constructions since the set of ``master
equations'' that need to be satisfied is now somewhat simpler than was
thought in the past.

Finally, we should mention two immediate extensions of the results in
this paper, both in supersymmetric directions.  One is to consider
also $\gh$ a semisimple Lie superalgebra {\sl with} nondegenerate
Cartan-Killing form.  The other extension keeps $\gh$ a semisimple Lie
algebra, but considers instead of $\ghhat$, the associated Ka{\v
c}-Todorov algebra.  This will be reported in \[FiMo].

\ack

It is pleasure to thank the Instituut voor Theoretische Fysica of the
Universiteit Leuven once more for their hospitality during the final
stages of this work.  Thanks also go individually to Alex Deckmyn,
Takashi Kimura, Noureddine Mohammedi, Niels Obers, Alexander Sevrin,
Sonia Stanciu, Kris Thielemans, and Walter Troost for their comments
and discussions.

\NoteAdded

After submission of this paper, Ezra Getzler informed me that this
construction is a special case of a method by which one may assign an
$N{=}2$ SCA to any Manin triple \[getzler].  Getzler's construction,
which seems to be implicit in \[leuven], has as two extreme special
cases the Kazama-Suzuki models on the one hand, and the present
construction ($G/G$) on the other.  I am grateful to him for his
correspondence and for sending me his preprint \[getzler] where, in
addition, one can also find the results of our Section 3.
\refsout
\bye